\documentclass[aps,amsfonts,nofootinbib,superscriptaddress,preprint]{revtex4-1}
\usepackage[normalem]{ulem} 
\makeatletter
\makeatletter
\usepackage{tikz}
\usetikzlibrary{decorations.pathmorphing, shapes.misc, shapes}
\usepackage{amsmath}
\usepackage{mdframed} 
\usepackage{tikz-feynman}
\tikzfeynmanset{compat=1.1.0}
\usepackage{tikz-feynman} 
\tikzfeynmanset{compat=1.1.0} 
\usepackage{feynmp-auto}
\usepackage{cleveref}
\usepackage{amssymb}
\usepackage{xcolor} 
\usepackage{tcolorbox} 
\usepackage{amsbsy}
\usepackage{bm}
\usepackage{epsfig} 
\usepackage{color}
\usepackage{slashed}
\usepackage{soul}
\usepackage{comment}
\usepackage{auncial}
\makeatletter
\usepackage{appendix}
\usepackage{epsfig} 
\usepackage{ulem}
\usepackage{pgfplots}
\newcommand{\stkout}[1]{\ifmmode\text{\sout{\ensuremath{#1}}}\else\sout{#1}\fi}
\newcommand{\ee}{\end{equation}}
\newcommand{\bb}{\begin{equation}}
\newcommand{\eqb}{\begin{eqnarray}}

\newcommand{\eqf}{\end{eqnarray}}

\usepackage[T1]{fontenc}
\usepackage{comment}
\usepackage[normalem]{ulem} 
\makeatletter
\usepackage{xcolor}
\usepackage{mdframed}
\usepackage{amsmath}
\usepackage{amssymb}
\usepackage{appendix}
\usepackage{amsbsy}
\usepackage{tikz}
\usepackage{booktabs}
\usetikzlibrary{decorations.pathmorphing, shapes.misc, shapes}
\usepackage{amsmath}
\usepackage{tikz}
\usepackage{feynmp-auto}
\usepackage{cleveref}
\usepackage{bm}
\usepackage{epsfig} 
\usepackage{color}
\usepackage{slashed}
\usepackage{soul}
\usepackage{comment}
\usepackage[normalem]{ulem} 
\makeatletter
\usepackage{amsmath}
\usepackage{amssymb}
\usepackage{appendix}
\usepackage{amsbsy}
\usepackage{bm}
\usepackage{epsfig} 
\usepackage{color}
\usepackage{slashed}
\usepackage{soul}
\begin{document}
\title{Entanglement and Effective Field Theories }
\author{Jorge Gamboa}
\email{jorge.gamboa@usach.cl}
\affiliation{Departamento de F\'{\i}sica, Universidad de Santiago de Chile, Santiago 9170020, Chile}

\begin{abstract}
We investigate the emergence of geometric phases in chiral transformations within gauge theories
coupled to fermions. We begin by analyzing the Schwinger model in (1+1) dimensions, where chiral
symmetry is explicitly modified due to the dynamical generation of a photon mass. This model
provides a controlled setting to study the interplay between anomalies and vacuum structure.

Building on these insights, we extend our analysis to four-dimensional QED by promoting the
vacuum angle $\theta$ to a dynamical field $\theta(x)$. This generalization allows us to explore how the axial
anomaly and the presence of a nontrivial vacuum structure modify the conventional chiral symmetry.

Using the adiabatic approximation, we demonstrate that chiral transformations are modified by
the emergence of a nontrivial Berry phase, which introduces a geometric correction that depends on
the topological properties of the vacuum. This result suggests that chiral transformations acquire
an effective gauge structure in parameter space, in the presence of a dynamical $\theta(x)$ field, leading
to new physical consequences at low energies.

This framework establishes a novel connection between chiral symmetries, anomalies, and geometric
phases, offering a unified approach to describing topological effects, vacuum structure, and
infrared modifications in gauge theories with fermions. Moreover, our results suggest that Berry
phases play a crucial role in the infrared structure of QED, potentially providing a mechanism for
regularizing infrared divergences in theories with axial anomalies.
\end{abstract}
\maketitle

\author{}
\date{}

\maketitle
\section{Introduction}
A fundamental cornerstone in developing computational methods in modern theoretical physics is the discovery of effective theories \cite{1,2,3}. The underlying idea behind the description of effective theories is that if we have the partition function:
\[\mathcal{Z} = \int \mathcal{D} \phi \, e^{-S[\phi]},\]   
and there exists an energy scale \(E \ll \Lambda\), where \(\Lambda\) is an ultraviolet (UV) cutoff, then integrating out the high-energy modes \(\phi_H\) results in:
\bb
\mathcal{Z} = \int \mathcal{D} \phi_L \, e^{-S_{\text{eff}}[\phi_L]},\label{1}
\ee
where \(S_{\text{eff}}\) is the effective action. 

 Generally, this process can be highly non-trivial because it requires developing sophisticated concepts associated with the renormalization group, effective coupling constants, and the renormalization flow \cite{1,3}. Significant results require considerable effort to achieve. However, fortunately, many of these ideas are well-documented in review papers \cite{kaplan,joe,shankar} and modern textbooks \cite{4}.

What we aim to discuss here, however, is a parallel development that is arguably as old as quantum mechanics itself. Specifically, we wish to explore the conceptual connection between the construction of entangled states and the framework of effective field theories. This comparison highlights the deep and complementary roles these ideas play in capturing the structure and dynamics of physical systems.

Quantum entanglement is a phenomenon in which two or more quantum systems are described by a joint state that cannot be factorized into independent states of each subsystem, even when spatially separated. This non-factorizability leads to strong quantum correlations, such that a measurement of one subsystem instantaneously influences the state of the other, regardless of the distance between them \cite{summers}. 
{{This is an aspect that has been addressed in the context of $C^*$-algebras and violations of Bell inequalities in the work of Summers and Werner, and has also been further developed in recent investigations, see for example \cite{varios12}.}}

The entanglement entropy precisely quantifies the amount of quantum correlation between subsystems \cite{nishi,cassini}. It is defined as the von Neumann entropy of the reduced density matrix of a given subsystem, obtained by tracing out the degrees of freedom of the complementary subsystem. This entropy measures the loss of information about the global system when only one part is observed. A zero value indicates a separable (unentangled) state, while a non-zero value signifies the presence of entanglement, with larger values corresponding to stronger quantum correlations.

The last paragraphs suggest that the connection between effective field theories and entanglement should be natural, as both frameworks aim to capture the essential degrees of freedom relevant at different scales. In this context, calculating entanglement entropy should emerge as a systematic procedure grounded in the same principles that govern the construction of effective actions. Just as integrating out high-energy modes leads to an effective description of low-energy physics, tracing out degrees of freedom in a quantum system provides a reduced description that inherently encodes entanglement.

A fully general treatment of these issues is an extremely challenging task. Therefore, a more appropriate approach is first to analyze a simple model and then attempt to generalize the results, if possible.  

One of the simplest quantum field theory models is the Schwinger model, which admits an exact solution and exhibits intriguing non-perturbative phenomena, such as the axial anomaly and, potentially, a $\theta$-vacuum.

 In this letter, we analyze this model using the adiabatic approximation and illustrate how its effective description offers a significant framework for understanding the ``low-energy regime'', our primary focus.
\section{Schwinger Model and its topological Implications}

This section will solve the Schwinger model \cite{schwinger,fidel} in the adiabatic approximation and analyze how the Berry phase emerges in the system's evolution \cite{berry}. In particular, we will show that the chiral charge and the Berry phase are intrinsically related and how this connection manifests through the chiral anomaly. We will demonstrate that the mass generation for the photon in 1+1 dimensions is a dynamical effect and reflects the presence of topological structures encoded in the Berry phase.

The partition function is   
\bb
Z= \int {\cal D} A {\cal D}{\bar \psi}{\cal D}\psi \,e^{-S}, \label{sch1}
\ee
where 
\bb
S = \int d^2x \left( \frac{1}{4} F_{\mu\nu} F^{\mu\nu} + \bar{\psi}( \gamma^\mu D_\mu [A]) \psi \right), \label{sch2}
\ee
and $D_\mu [A]= \partial_\mu + ie A_\mu$.

The functional measure is not invariant under local chiral transformations
\bb
\psi (x) \to  e^{i \alpha (x) \gamma_5}~ \psi, ~~~~~{\bar \psi} (x) \to  {\bar \psi} ~e^{i \alpha (x) \gamma_5}. \label{chiral}
\ee

 Therefore, to have a properly defined measure, we must perform the appropriate substitution \cite{fujikawa}
\bb
{\cal D}{\bar \psi}{\cal D}\psi \to {\cal D}{\bar \psi}{\cal D}\psi \,e^{-\frac{e}{\pi} \int d^2x \,\alpha\,{\tilde F}}. \nonumber
\ee

Thus, the partition function takes the form  
\bb
Z= \int {\cal D} A\, {\cal D}{\bar \psi}{\cal D}\psi \, \,e^{-\frac{e}{\pi} \int d^2x \,\alpha\,{\tilde F}}\,e^{-\int d^2 \left[\frac{1}{4} F^2 + {\bar \psi} \left(\slashed D [A] -i\partial_\mu \alpha \gamma_\mu \gamma_5\right)\psi  \right]}. \label{sch4}
\ee

Now, let us solve the eigenvalue problem in the adiabatic approximation
\bb
\left({\slashed D} [A] - i\partial_\mu \alpha \gamma^\mu \gamma^5\right) \varphi_n = \lambda_n \varphi_n, \label{sch5}
\ee

In $1+1$ dimensions, we use the  Dirac matrices representation
\[
\gamma^0 = \sigma^1, \quad \gamma^1 = \sigma^2, \quad \gamma^5 = \gamma^0 \gamma^1 = -\sigma^3.
\]

In the adiabatic approximation, where \(\alpha(x)\) varies slowly in spacetime, we write the solution in terms of the Berry phase $\mathcal{A}(\alpha)$
\begin{equation}
\varphi_n(x) = e^{i \gamma^5 \gamma_n(x) }\tilde{\varphi}_n(x),
\end{equation}
where $\gamma_n(x)$ is the geometric phase and ${\tilde \phi}_n (x)$ a wave function to be determined.

Substituting this into the eigenvalues equation, we obtain:

\begin{equation}
\left( {\slashed D} [A] - i\partial_\mu \alpha \gamma^\mu \gamma^5 + i \gamma_5 \slashed{\partial} \gamma_n(x) \right) \phi_n(x) = \lambda_n \phi_n(x).
\end{equation}

For consistency with the adiabatic approximation, the new term $i \gamma_5 \slashed{\partial} \gamma_n(x)$ must compensate for $i\partial_\mu \alpha \gamma^\mu \gamma^5$, leading to the equation:
\begin{equation}
\slashed{\partial} \gamma_n(x) = \partial_\mu \alpha \gamma^\mu.
\end{equation}
which implies: 
\begin{equation}
\gamma_n (x) =  \alpha (x). \label{chibe}
\end{equation}

With this choice, the equation for ${\tilde \varphi}_n(x)$ simplifies to:
\begin{equation}
{\slashed D} [A] {\tilde \varphi}_n = \lambda_n {\tilde \varphi}_n.
\end{equation}
which is just the Dirac equation without the term $\partial_\mu \alpha \gamma^\mu \gamma^5$, confirming that the Berry phase has absorbed this term.
This gives us the accumulated Berry phase:

From this result, we conclude for the Schwinger model that:

\begin{enumerate}
    \item The Berry phase is exactly the same as the local chiral transformation.
    \item The system effectively interprets the Berry phase as a geometric connection in parameter space, closely related to the axial anomaly.
\end{enumerate}

Since $\gamma^5$ generates chiral transformations, this result indicates that the Berry phase can be interpreted as a chiral gauge transformation. This is directly related to the chiral anomaly \cite{adler,bell} in the Schwinger model, reflecting its topological structure.
Let us analyze the relationship between the chiral charge and the Berry phase. The chiral anomaly implies $\partial_\mu j^\mu_5 = \frac{e}{\pi} \epsilon_{\mu \nu} F_{\mu \nu}$ and from this equation, we obtain
\eqb 
\frac{d Q_5}{dt} &=& \frac{e}{\pi} \int dx \, \epsilon_{\mu \nu} F_{\mu \nu} \nonumber
\\
&=& \frac{e}{\pi} \int dx \, E(x), \label{schw1}
\eqf
where $E(x)$ is te electric field.

The Berry phase in the eigenvalue problem was obtained in the adiabatic approximation through the transformation 

\begin{equation}
\varphi_n(x) = e^{i \gamma^5 \alpha(x)} \tilde{\varphi}_n(x).
\end{equation}

The accumulated Berry phase over a path in parameter space \(\alpha\) is given by the integral of the Berry connection \(\mathcal{A}(\alpha)\):

\begin{equation}
\gamma_B = \oint d\alpha \, \mathcal{A}(\alpha).
\end{equation}

Since we found that the Berry phase is \(\mathcal{A}(\alpha) = \alpha\), we obtain:

\begin{equation}
\gamma_B = \oint d\alpha \, \alpha = \alpha_{\text{final}} - \alpha_{\text{initial}}.
\end{equation}

If \(\alpha(x)\) parametrizes a local chiral transformation, we can connect this result to the chiral charge.

The local chiral transformation is given by:

\begin{equation}
\psi \to e^{i \gamma^5 \alpha(x)} \psi. \label{chi}
\end{equation}

Comparing this with the solution obtained in the eigenvalue problem, we see that the Berry phase acts as a local chiral transformation. Now, integrating the chiral anomaly equation over time gives:

\begin{equation}
Q_5(t) - Q_5(0) = \frac{e}{\pi} \int d^2x \, E(x).
\end{equation}

This equation shows that the variation of the chiral charge is related to the presence of an electric field in \(1+1\) dimensions.

From the eigenvalue problem solution, the Berry phase \(\mathcal{A}(\alpha) = \alpha\) effectively represents the chiral transformation undergone by the system, and the variation of the chiral charge in time is obtained from the accumulated phase.

Thus, we can identify the variation of the chiral charge with the accumulated Berry phase:

\begin{equation}
\gamma_B = \Delta Q_5. \label{chiral}
\end{equation}

The difference $\Delta Q_5$ can be computed using the Atiyah-Singer theorem, given by 
\bb
\mbox{Index} (\slashed D) = -\frac{e}{\pi} \int d^2x {\tilde F} = n_+-n_-= \gamma_B, 
\ee 
where $n_\pm$ represents the zero modes of the Dirac operator with positive and negative chiralities, respectively.

Equation (\ref{chiral}) states that the Berry phase accumulated in the adiabatic approximation is an integer, reflecting its topological origin. The Atiyah-Singer theorem \cite{atiyah} directly relates the anomaly to the Dirac operator, establishing a deep connection between gauge space topology, the spectral structure of fermions, and the geometric evolution of the system

\section{QED in Four Dimensions}

The extension of gauge theory to four dimensions does not follow directly from a naive extension of the Schwinger model. Instead, it arises by considering the vacuum parameter $\theta$ as an adiabatic parameter. Indeed, instead of writing the partition function in the standard form, we take the modified expression:

\begin{equation}
Z = \int {\cal D} A_\mu {\cal  D} {\bar \psi} {\cal D} \psi\, J(\alpha)\,e^{-S},
\end{equation}
\noindent 
where \( J(\alpha) \) is the Fujikawa Jacobian, which captures the non-invariance of the fermionic measure under a local chiral transformation, and its explicit form is
\begin{equation}
J(\alpha) = e^{- \frac{g^2}{16 \pi^2} \int d^4 x \,\alpha(x) F_{\mu \nu} {\tilde F}^{\mu \nu}}.
\end{equation}

The action in Euclidean space is

\begin{equation}
S = \int d^4x \left( \frac{1}{4} F_{\mu\nu} F_{\mu\nu} + \bar{\psi} \left( \slashed{D}[A] - i \slashed \partial {\alpha} \gamma_5 \right)\psi +\frac{g^2}{16 \pi^2} \alpha(x) F_{\mu \nu} {\tilde F}_{\mu \nu} + \frac{\theta}{16 \pi^2}F_{\mu \nu} {\tilde F}_{\mu \nu} \right). \label{effe1}
\end{equation}
and 
\[
\tilde{F}_{\mu\nu} = \frac{1}{2} \epsilon_{\mu\nu\rho\sigma} F_{\rho\sigma}.
\]

The effective action (\ref{effe1}) contains corrections due to chiral symmetry and the fermionic measure. The term $\theta$ is present in the theory from the beginning as a vacuum angle, but in the adiabatic approximation, we promote it from a constant parameter to a spacetime-dependent field, that is, $\theta \to \theta(x)$. This allows us to compute the fermionic determinant in the adiabatic approximation, capturing both the axial anomaly and the geometric effects associated with the Berry phase.

{{We compute the fermionic determinant, solving the eigenvalue equation
\bb
\left( \slashed{D}[A]  + \slashed{\cal A}^n- i\slashed \partial {\alpha} \gamma_5 \right)\varphi_n = \lambda_n \varphi_n.
\ee}}

To solve this equation, we use
\bb
\varphi_n (x) = e^{i \gamma_5\gamma_n (x)}\phi_n (x),
\ee
where $\gamma_n$ is the Berry's phase \cite{comment}.

Taking into account the above considerations, (\ref{effe1}) becomes
\begin{equation}
S = \int d^4x \left( \frac{1}{4} F_{\mu\nu} F^{\mu\nu} + \bar{\psi} \left( \slashed D[A] - \gamma_5\slashed  {\cal A}^n - i \slashed \partial {\alpha} \gamma_5 \right)\psi+  \frac{1}{16 \pi^2} \left(-g^2 \alpha(x) + \theta (x)\right) F_{\mu \nu} {\tilde F}^{\mu \nu} \right),  \label{effe2}
\end{equation}
where 
\bb
{\cal A}_\mu^n = \partial_\mu \gamma_n (x)= i \langle \varphi_n|\partial_\mu |\varphi_n\rangle. \label{berr}
\ee 
is the Berry connection.

Following the analogy with the Schwinger model, we choose
\bb
\gamma^n = \alpha
\ee
where $\gamma^n$ denotes the Berry phase accumulated in the $n$-th mode. 

{{Then, the effective $\theta$ angle, appearing in the term proportional to 
$
\tilde{F}_{\mu \nu}F^{\mu \nu}$, 
becomes
\bb
\theta_\text{eff} = -g^2 \gamma^n + \theta.
\ee}}
{{This expression relates the effective $\theta$ angle to the Berry phase, just as in the Schwinger model the Berry phase is directly linked to the anomalous contribution to the effective action.}}

We obtain the final effective action 
\begin{equation}
S_{\mbox{eff}} = \int d^4x \left( \frac{1}{4} F_{\mu\nu} F^{\mu\nu} + \bar{\psi}  \slashed D \psi +  \theta (x) \left[ \partial_\mu \left(J_5^\mu + \bar{\psi} \gamma_5 \gamma^\mu \partial_\mu \gamma_n \psi \right) - \frac{g^2}{16\pi^2} F_{\mu\nu} \tilde{F}^{\mu\nu}\right] \right). \label{effe3}
\end{equation}

{{From these equations, we observe that the chiral anomaly can be reinterpreted in terms of the Berry phase, which emerges as a low-energy effect. In particular, the usual chiral transformation, from the perspective of the adiabatic approximation, admits a topological interpretation similar to a holonomy. }}

Mathematically, this implies that the effective chiral transformation in the low-energy regime cannot be described solely by a conventional rotation in spinor space but must incorporate the Berry connection $\mathcal{A}_\mu^n$ 
in the relevant parameter space. Consequently, the covariant derivative of the chiral current acquires an additional term reflecting the Berry curvature, given by
\[
\mathcal{F}_{\mu\nu} = \partial_\mu \mathcal{A}_\nu - \partial_\nu \mathcal{A}_\mu.
\]
This result suggests that the parallel transport of quantum states in parameter space is nontrivial, which modifies the interpretation of chiral symmetry at the effective level.

Physically, this modification implies that chiral symmetry ceases to be a well-defined global symmetry in the effective regime, as the Berry phase introduces a geometric dependence that may induce topological effects in the theory. Furthermore, the chiral anomaly, traditionally interpreted as a consequence of the non-invariance of the functional measure under chiral transformations, could acquire an alternative interpretation in terms of the curvature of the low-energy state bundle. This suggests that the nature of the anomaly could be intrinsically linked to the geometry of the parameter space, providing a novel perspective on its origin and manifestation in effective field theories.

\section{Are Entanglement and Berry phases connected Through Anomalies?.}

In this section, we will answer the previously posed question and explain how entanglement
entropy is related to the chiral anomaly in the discussed examples. A priori, this
connection is not trivial, since, although entanglement entropy is consistent with the philosophy
of effective field theories, anomalies introduce subtle effects that require a detailed
explanation.
To discuss our ideas again, using the Schwinger model as an initial ingredient and calculating
the entanglement entropy is illustrative. 

For this, let us write the matrix density path integral representation:
\bb
\rho [A,A']= \int {\cal D}{\bar \psi} {\cal D}\psi \,e^{-S[A,{\bar \psi},\psi]}. \label{den1}
\ee
where $S$ is given (\ref{sch2}).

To obtain the reduced density matrix $\rho_A$ for a subregion $A$, we divide the spacetime into
two regions, $A$ and $B$, and trace over the degrees of freedom in $B$
\bb
\rho_A[A,A'] = \int_B {\cal D} A_B \rho[A,A'].
\ee

This path integral representation describes the entanglement structure of the vacuum
state by summing over all possible field configurations in region $B$ while keeping the boundary
conditions in region $A$ fixed at $A$ and $A'$.

Under a local chiral transformation (\ref{chi}), the fermionic measure transforms as in (\ref{chi}) but
with the identification (\ref{chibe}). The previous identifications lead to the entanglement entropy
for the Schwinger model being:
\bb
S_A = \frac{1}{6} \log \left(\frac{L}{\epsilon}\right) + \beta |\delta Q_5|,
\ee
where the entanglement entropy $ \frac{1}{6} \log \left(\frac{L}{\epsilon}\right)$ receives a correction from the Berry phase contribution.
The constant $\beta$ can be computed using the Atiyah-Singer theorem and has the
form $\log \frac{1}{\epsilon}$. However, this divergence is regularized by the photon mass, which is dynamically
generated in the Schwinger model.
The extension to four dimensions is more complicated—although physically very interesting—
and requires more or less the same identifications that we have considered in the
Schwinger model, in this case, are the equations (\ref{effe2}) and (\ref{effe3}).

The final result for the entanglement entropy is\footnote{Note that (\ref{entropy}) is another example of topological entropy. although different from the one discussed in \cite{preskill,otro}.}
\bb
S_A = S_0 + \beta \int_\Sigma d^3 x \theta(x) \left( \frac{g^2}{16 \pi^2} F_{\mu \nu} {\tilde F}^{\mu \nu} + {\bar \psi} \gamma_5 \gamma^\mu \partial_\mu \gamma_n \psi \right), \label{entropy}
\ee
where $\Sigma$ is the entanglement surface and $\beta$ is a coefficient related to the dynamically generated
mass scale in the theory and takes the form
\bb 
\beta \sim \ln \frac{1}{m_{gap}},
\ee
and $m_{gap}$ is the effective energy scale generated by the Berry phase and the axial anomaly.
In conclusion, entanglement entropy and the Berry phase appear connected here through
the chiral anomaly. To our knowledge, this effect has not been noticed before.
\section{Final Comments}

The mass $m_{gap}$ is nonzero and can be obtained from the analysis of the equations of motion
and the perturbative calculation of the fermionic determinant, from which one derives:
\bb
{\cal L}_{eff} \supset -\frac{1}{2} f_n^2  (\partial_\mu \gamma_n) (\partial^\mu \gamma_n) + \frac{g^2}{16 \pi^2} \gamma_n F_{\mu \nu}{\tilde F}^{\mu \nu}.
\ee

This result is very interesting because the appearance of an axion-like coupling in QED
can modify the infrared structure of the theory as a consequence of the anomaly, potentially
regularizing IR divergences in certain regimes. This introduces a geometric and topological
interpretation to the long-standing problem of soft photons in QED. From this perspective,
one possible answer to how IR divergences in QED are avoided is through the emergence of
Berry phases at low energies. However, whether this mechanism fully resolves all IR issues
in QED remains an open question.

This work is dedicated to the memory of my friend, Víctor O. Rivelles, whose guidance and insights profoundly influenced my early scientific work. It is a pleasure to thank Horacio Falomir for the discussions. Dicyt (USACH) 042531GR$\_$REG
supported this research.


\begin{thebibliography}{99}
\bibitem{1} K.~G.~Wilson and J.~B.~Kogut,
Phys. Rept. \textbf{12} (1974), 75-199
doi:10.1016/0370-1573(74)90023-4.
\bibitem{2} S.~Weinberg,
Physica A \textbf{96} (1979) no.1-2, 327-340
doi:10.1016/0378-4371(79)90223-1: ibid, S.~Weinberg,
Phys. Lett. B \textbf{91} (1980), 51-55
doi:10.1016/0370-2693(80)90660-7.

\bibitem{3}
J.~Berges, N.~Tetradis and C.~Wetterich,
Phys. Rept. \textbf{363} (2002), 223-386
doi:10.1016/S0370-1573(01)00098-9
[arXiv:hep-ph/0005122 [hep-ph]].
\bibitem{kaplan} D. B. Kaplan, [arXiv:nucl-th/0510023 [nucl-th]].
\bibitem{joe} J. Polchinski, [arXiv:hep-th/9210046 [hep-th]].
\bibitem{shankar}R. Shankar, Rev. Mod. Phys. 66 (1994), 129-192 doi:10.1103/RevModPhys.66.129 [arXiv:condmat/
9307009 [cond-mat]].
\bibitem{4} C.~P.~Burgess,
Cambridge University Press, 2020,
ISBN 978-1-139-04804-0, 978-0-521-19547-8
doi:10.1017/9781139048040.
\bibitem{summers} S. J. Summers and R. Werner, J. Math. Phys. 28, 2448 (1987); doi: 10.1063/1.527734.
\bibitem{varios12} M.~S.~Guimaraes, I.~Roditi, S.~P.~Sorella and A.~F.~Vieira, ``Relative entropy of squeezed states in Quantum Field Theory,'', arXiv:2504.13148 [hep-th];  R.~Grossi and J.~C.~A.~Barata,``A categorical view of Bell's inequalities in quantum field theory'', arXiv:2409.14930 [math-ph]; P.~De Fabritiis, M.~S.~Guimaraes, I.~Roditi and S.~P.~Sorella,``Numerical approach to the Bell-Clauser-Horne-Shimony-Holt inequality in quantum field theory'', Phys. Rev. D \textbf{110} (2024) no.6, 065006.

\bibitem{nishi} 
T.~Nishioka, S.~Ryu and T.~Takayanagi,
J. Phys. A \textbf{42} (2009), 504008
doi:10.1088/1751-8113/42/50/504008
[arXiv:0905.0932 [hep-th]]; T.~Nishioka,
Rev. Mod. Phys. \textbf{90} (2018) no.3, 035007
doi:10.1103/RevModPhys.90.035007
[arXiv:1801.10352 [hep-th]].

\bibitem{cassini} H.~Casini, M.~Huerta and R.~C.~Myers,
JHEP \textbf{05} (2011), 036
doi:10.1007/JHEP05(2011)036
[arXiv:1102.0440 [hep-th]]; ibid, 
J. Phys. A \textbf{42} (2009), 504007
doi:10.1088/1751-8113/42/50/504007
[arXiv:0905.2562 [hep-th]].

\bibitem{schwinger} J. S. Schwinger, Phys. Rev. 125 (1962), 397-398 doi:10.1103/PhysRev.125.397.

\bibitem{fidel} R. E. Gamboa Saravi, M. A. Muschietti, F. A. Schaposnik and J. E. Solomin, Annals Phys. 157
(1984), 360 doi:10.1016/0003-4916(84)90065-4; R. E. Gamboa Saravi, F. A. Schaposnik and
J. E. Solomin, Nucl. Phys. B 185 (1981), 239-253 doi:10.1016/0550-3213(81)90375-8; P. Arias,
H. Falomir, J. Gamboa, F. Mendez and F. A. Schaposnik, Phys. Rev. D 76 (2007), 025019
doi:10.1103/PhysRevD.76.025019 [arXiv:0705.3263 [hep-th]].




\bibitem{berry} M.~V.~Berry,
Proc. Roy. Soc. Lond. A \textbf{392} (1984), 45-57
doi:10.1098/rspa.1984.0023.
\bibitem{fujikawa} K.~Fujikawa,
Phys. Rev. D \textbf{21} (1980), 2848
[erratum: Phys. Rev. D \textbf{22} (1980), 1499]
doi:10.1103/PhysRevD.21.2848 and 
Phys. Rev. Lett. \textbf{42} (1979), 1195-1198
doi:10.1103/PhysRevLett.42.1195.

\bibitem{adler} S. L. Adler, Phys. Rev. 177 (1969), 2426-2438 doi:10.1103/PhysRev.177.2426; S. L. Adler and
W. A. Bardeen, Phys. Rev. 182 (1969), 1517-1536 doi:10.1103/PhysRev.182.1517.

\bibitem{bell} J. S. Bell and R. Jackiw, Nuovo Cim. A 60 (1969), 47-61 doi:10.1007/BF02823296.


\bibitem{atiyah} M. F. Atiyah, I. M. Singer. Bulletin of the American Mathematical Society, 69(3), 422-433
(1968). DOI: 10.1090/S0002-9904-1963-10957-X; L. Alvarez-Gaume, L., and P. Ginsparg, P.
11 Annals of Physics, 161(2) (1984), 423-490. DOI: 10.1016/0003-4916(85)90087-5.


\bibitem{comment}Rigorously, one should also consider the dynamical phase $\lambda_n$, which depends on the spectrum of the Dirac operator. The connection of Berry $\mathcal{A}_\mu$ is computed from the differences of eigenvalues $\lambda_n - \lambda_m$, meaning their structure remains crucial even in the low-energy regime. However, when topological effects dominate, the absolute energy levels of $\lambda_n$ are less relevant than their differences. If these differences remain small but nonzero, the calculation of $\mathcal{A}_\mu$ remains well-defined, and the contribution of the dynamical phase to the effective action is subleading. Therefore, we do not set $\lambda_n = 0$ exactly, but we assume it plays a minor role compared to topological effects.

\bibitem{preskill} A. Kitaev and J. Preskill, Phys. Rev. Lett. 96 (2006), 110404
doi:10.1103/PhysRevLett.96.110404 [arXiv:hep-th/0510092 [hep-th]].

\bibitem{otro} For applications and references see: M. Barkeshli, P. Bonderson, M. Cheng and Z. Wang,
Phys. Rev. B 100 (2019) no.11, 115147 doi:10.1103/PhysRevB.100.115147 [arXiv:1410.4540
[cond-mat.str-el]].




 \end{thebibliography}
\end{document}